\documentclass[12pt,preprint]{aastex}

\shorttitle{Behavior of X-Ray Dust Scattering} \shortauthors{Shao &
Dai}

\begin{document}
\title{Behavior of X-Ray Dust Scattering and Implications for X-Ray Afterglows of Gamma-Ray Bursts}

\author{L. Shao and Z. G. Dai}
\affil{Department of Astronomy, Nanjing University, Nanjing 210093,
China \\ lang@nju.edu.cn, dzg@nju.edu.cn}

\begin{abstract}
The afterglows of gamma-ray bursts (GRBs) have commonly been assumed
to be due to shocks sweeping up the circum-stellar medium. However,
most GRBs have been found in dense star-forming regions where a
significant fraction of the prompt X-ray emission can be scattered
by dust grains. Here we revisit the behavior of dust scattering of
X-rays in GRBs. We find that the features of some X-ray afterglows
from minutes to days after the gamma-ray triggers are consistent
with the scattering of prompt X-ray emission from GRBs off host dust
grains. This implies that some of the observed X-ray afterglows
(especially those without sharp rising and decaying flares) could be
understood with a dust-scattering--driven emission model.
\end{abstract}

\section{INTRODUCTION}

Currently, the popular model for gamma-ray bursts (GRBs) and their
afterglows is the fireball-shock model (for recent reviews, see
Zhang \& M\'esz\'aros 2004; Piran 2005; M\'esz\'aros 2006). In this
model, the short-term prompt emission of GRBs is ascribed to
internal shocks in the ejecta, and the long-term afterglow at lower
energy is ascribed to external shocks sweeping up the circum-stellar
medium. The rapid localization of the {\em Swift Gamma-Ray Burst
Explorer} \citep{g04} has led to a recent breakthrough in the
detections of early afterglows. This has given rise to theoretical
studies of the early afterglow emission, especially in X-ray energy,
thanks to the X-Ray Telescope (XRT) onboard {\em Swift}.

Recently, an X-ray halo around the short GRB 050724 was detected by
XRT, with both a radial temporal evolution and intensity
distribution that are consistent with the properties of Galactic
dust-scattering \citep{v06}. Previously, GRB 031203 \citep{v04} and
GRB 050713A \citep{tm06} had also been found to have similar
dust-scattered X-ray halos, as observed by {\em XMM-Newton}. In
addition, a careful inspection of the {\em XMM-Newton} data shows
the presence of some diffuse emission in the field of GRB 050730,
which could also be produced by dust-scattering, although this needs
to be verified \citep{tm06}. Note that host galaxies have been found
for some GRBs localized so far. Most of these galaxies show signs of
active star formation, implying the presence of GRB progenitors
forming out of dense gaseous clouds \citep{p98}. Thus,
dust-scattering off of dust grains may be common in GRB phenomena,
which could have been playing an important role in the observed
X-ray afterglows.

A diffuse X-ray halo is predicted to appear around an X-ray point
source when the interstellar dust grains scatter some of the X-rays,
typically by $1\arcmin$ to 1$^\circ$ \citep{o65,m70}. This
time-dependent information about scattering in GRBs was previously
considered by \cite{d91}, who assumed the existence of binary
companions or accretion disks in GRB systems; a Compton echo of
reflected X-ray and gamma-ray emission with a time profile mimicking
the primary burst emission was expected. The features of delayed
echo emission were also discussed in detail by \cite{m99},
\cite{mg00}, \cite{m00}, \cite{eb00}, \cite{ss03} and \cite{rm04} in
a variety of emission geometries and ambient gas distributions
around GRBs.

In this paper, the scattering of X-rays by dust grains in GRBs is
revisited. We estimate the emerging flux during such an X-ray echo
event, which is expected to be dominant in the X-ray afterglow. We
find that an initial pulse of X-rays from a normal GRB scattering
off dust grains in a host galaxy can give rise to a long-term
``afterglow" with almost the same amount of energy as expected in
GRB X-ray afterglows. However, the angular size of this echo
emission is too small to be possibly resolved on Earth. Therefore,
only the temporal features of the total scattered flux are
considered here. We find that some of these features are consistent
with observations of GRB X-ray afterglows. We discuss the prominent
flux of X-ray echo emission in $\S$ 2 and the temporal behavior in
$\S$ 3. We suggest in $\S$ 4 that dust scattering of prompt X-ray
emission off host dust grains may be an alternative explanation for
some of the GRB X-ray afterglows. Our conclusions are summarized in
$\S$ 5.

\section{TOTAL FLUX OF AN X-RAY ECHO}
The quantity that we first consider is the amount of energy. This
should be evaluated for the scattering effect in the context of GRBs
at the first step to make sure that scattering off dust grains can
viably produce a detectable X-ray afterglow. This was previously
discussed by \cite{m99}. We follow his result here.

Considering a variable X-ray source at angular diameter distance
$D_s$, and an intervening dust layer (e.g. a galaxy or a cloud) at
redshift $z_d$ and with angular diameter distance $D_d$, the time
delay $t_d$ of photons received at an angle $\theta$ from the source
scattered by the dust layer is given by
\begin{equation}
t_d={(1+z_d)D_d D_s \theta^2 \over{2 c D_{ds}}}\,,
\end{equation}
where $D_{ds}$ is the angular diameter distance from the dust layer
to the source. See Figure 1 for the geometry of the scattering.

For simplicity, the variable source is assumed to emit most of its
energy in a narrow energy band (e.g. 2--10 keV) with a given fluence
$S_0$, in a time much shorter than $t_d$. Thus we can approximate
the source emission as a pulse of monochromatic light with
wavelength $\lambda$. If an experiential size distribution of grains
in the intervening dust layer is assumed, i.e., $\tau(a)\propto
a^{-0.5}$, where $\tau(a)$ is the scattering optical depth of dust
grains with radius $a$, the total flux of the X-ray echo due to the
small-angle scattering can be estimated by
\begin{equation}
F_h\simeq1.96\times10^{-7}{S_0 D_s \theta a_p^{-0.5}\over{D_{ds}
t_d \lambda }}{\bar{N_b}\over{10^{22}\,{\rm
cm^{-2}}}}{Z_t\over{0.02}}\,,
\end{equation}
where $a_p\simeq1.5\lambda D_{ds}/(2\pi D_s \theta)$ is
approximately satisfied, $\bar{N_b}$ is the mean gas column density
over the layer, and $Z_t$ is the total metallicity in dust grains.
Here the differential cross section is treated as an approximation
of the Mie solution, which is given in the form of the
``Rayleigh-Gans" approximation in $\S$ 3.

Using the relation between $t_d$ and $\theta$ (eq. [1]) in equation
(2) yields
\begin{equation}
F_h(t_d)\simeq48.5\left((1+z_d)D_s\over{D_d D_{ds}
}\right)^{3/4}{S_0 t_d^{-1/4}\over{\lambda^{3/2}}}
{\bar{N_b}\over{10^{22}\,{\rm cm^{-2}}}}{Z_t\over{0.02}}\,,
\end{equation}
where the echo flux is a function of the time delay $t_d$. We should
point out that (1) this equation is a good estimate of the flux of
X-ray echoes, from which we can get a preliminary idea of how much
the flux will be at a given $t_d$ and whether it will be detectable
at the time of interest. We discuss this issue later in this
section. (2) Considering the typical scattering angle implied by the
value of $a_p$, equation (3) is only valid before or around $t_d\sim
10^{10}\,{\rm s}(9\lambda^2/32\pi^2
c)(1+z_d)(D_{ds}D_d/D_s)(a_p/0.1\,\mu{\rm m})^{-2}$, where $c$ is
the speed of light. (3) This equation implies that a pulse of light
propagating through a dusty region will produce a long-term delayed
emission, which is like an ``afterglow" of this pulse (see Fig. 1).
Here we call this a dust-scattering--driven afterglow. We work out
its detailed light curve in $\S$ 3.

Equation (3) suggests that the flux of echoes is substantially
determined by the given geometry of dust scattering (i.e., $D_s$,
$D_d$, and $D_{ds}$) at a given time. In general, three cases should
be considered.

Case (1): the dust layer is in the form of an intervening galaxy,
which is also at a cosmological distance. Thus,
\begin{eqnarray}
F_h&=&1.27\times10^{-15} \left({S_0\over10^{-6}{\rm
ergs\,cm^{-2}}}\right) \left({\epsilon\over6 {\rm
keV}}\right)^{3/2}\left({t_d\over1{\rm yr}}\right)^{-1/4}
\nonumber \\ & & \times\ {\bar{N_b}\over{10^{22}\,{\rm
cm^{-2}}}}{Z_t\over{0.02}}\,{\rm ergs\,cm^{-2}\,s^{-1}}\,,
\end{eqnarray}
where $\epsilon$ is the mean energy of the X-ray photons,
$D_d/(1+z_d)\simeq10^3$ Mpc, and $D_s/D_{ds}\simeq2$ are assumed.
This X-ray echo can be much brighter than any intrinsic emission
from the intervening galaxy \citep{m99}. However it is below the XRT
detection limit ($\sim 2\times10^{-14}{\rm ergs\,cm^{-2}\,s^{-1}}$
for a $10^4$ s integration time) of {\em Swift} \citep{g04}.

Case (2): the dust scattering happens at low latitudes or in dense
clouds of our Galaxy, which have been verified by observations. We
can get a similar expression for the total flux of the echo
\begin{eqnarray}
F_h&=&1.34\times10^{-10} \left({S_0\over10^{-6}{\rm
ergs\,cm^{-2}}}\right) \left({\epsilon\over6 {\rm
keV}}\right)^{3/2}\left({t_d\over10^3{\rm s}}\right)^{-1/4}
\nonumber \\ & & \times\left({D_d\over100 {\rm
pc}}\right)^{-3/4}{\bar{N_b}\over{10^{21}\,{\rm
cm^{-2}}}}{Z_t\over{0.02}}\,{\rm ergs\,cm^{-2}\,s^{-1}}\,.
\end{eqnarray}
where $z_d=0$, $D_s\simeq D_{ds}$, $D_d\simeq 100$ pc,  and
$\bar{N_b}\simeq 10^{21}\,{\rm cm^{-2}}$ are assumed \citep{dl90}.
The typical halo radius is about $\theta\simeq 1.5'(t_d/10^3\,{\rm
s})^{1/2}(D_d/100\,{\rm pc})^{-1/2}$. This echo feature is quite
detectable and consistent with observations (Vaughan et al. 2004,
2006). Due to the large halo radius, this case can be easily
distinguished by observational analysis \citep{tm06}.

Case (3): if the dust scattering happens in the host galaxy of a GRB
at redshift $z_d\simeq1$, we have
\begin{eqnarray}
F_h&=&2.25\times10^{-9}\left({S_0\over10^{-6}{\rm
ergs\,cm^{-2}}}\right) \left({\epsilon\over6 {\rm
keV}}\right)^{3/2}\left({t_d\over10^3{\rm
s}}\right)^{-1/4}\nonumber \\ & & \times\left({D_{ds}\over 100
{\rm
pc}}\right)^{-3/4}\left({1+z_d}\over2\right)^{3/4}{\bar{N_b}\over{10^{22}\,{\rm
cm^{-2}}}}{Z_t\over{0.02}}\,{\rm ergs\,cm^{-2}\,s^{-1}}\,.
\end{eqnarray}
where $D_s\simeq D_{d}$, $D_{ds}\simeq 100$ pc,  and
$\bar{N_b}\simeq 10^{22}\,{\rm cm^{-2}}$ are assumed. This case is
similar to case (2), and thus the echo emission is significantly
detectable, since both cases have a similar scattering geometry,
except that the halo in case (3) has a very small angular size and
cannot be resolved on Earth. Obviously such a bright flux is quite
comparable to those of normal X-ray afterglows (e.g. Costa 1999). We
need to give a further consideration of this echo component, which
may have been observed but not realized so far. In general, only the
total flux of the echo emission versus time needs to be considered,
due to its small angular size.

\section{LIGHT CURVE OF AN X-RAY ECHO}

Here, we revisit the temporal behavior of an X-ray echo event with a
delayed time of minutes to days (e.g. M\'{e}sz\'{a}ros \& Gruzinov
2000; Sazonov \& Sunyaev 2003). For this purpose, we need to know
the differential cross section for the small-angle scattering of
X-rays off dust grains. For typical spherical grains of radius $a$,
in the limit of
\begin{equation}
\left({\epsilon\over1 {\rm keV}}\right)^{-1}\left({a\over1\mu{\rm
m}}\right)\ll1\,,
\end{equation}
the exact Mie solution recovers the Rayleigh-Gans approximation
\begin{equation}
{{\rm d}\sigma\over {\rm
d}\Omega_{SC}}=8\pi\sigma_T\left(a\over\lambda\right)^2{j_1^2(x)\over
x^2}\,,
\end{equation}
where $\sigma_T$ is the total cross section, $\lambda$ is the
wavelength of the X-ray photons, $x\equiv(2\pi a/\lambda)\alpha$ is
the scaled angular coordinate, $\alpha$ is the scattering angle, and
$j_1(x)=({\rm sin}\, x)/x^2-({\rm cos}\, x)/x$ is the first-order
spherical Bessel function \citep{o65,ah78,sd98}. For simplicity, our
treatment takes the dust grains as Rayleigh-Gans particles (e.g.
Kr\"{u}gel 2003) and is independent of the chemical composition or
shape of the grains, as long as equation (7) holds.

\subsection{Analytical Treatment in a Simple Case}

For a variable X-ray source (e.g. the prompt X-ray counterpart of a
GRB) at an angular diameter distance $D_s$, with an unabsorbed flux
as a function of time, $F_u(t)$, scattered by an intervening dust
layer (which could be assumed to be in the host galaxy of a GRB with
an angular diameter distance $D_{ds}$ from the GRB) at an angular
diameter distance $D_d$ with a scattering optical depth
$\tau(\theta,\phi)$, the observed intensity $I$ of the X-ray echo is
calculated by
\begin{equation}
I(\theta,\phi;t)=F_u(t-t_d)\tau(\theta,\phi){{\rm d}\sigma\over
\sigma_T {\rm d}\Omega_{SC}}{{\rm d}\Omega_{SC}\over {\rm d
}\Omega}={4aD_s F_u(t-t_d)\tau(\theta,\phi)j_1^2(x)\over \lambda
D_{ds}x\theta}\,,
\end{equation}
where $t_d$ is given by equation (1), $\alpha=(D_s/D_{ds})\theta$, $
d\Omega_{SC}=\alpha d\alpha d \phi$, and $d \Omega=\theta d\theta d
\phi$ are used in the small-angle limit, $\lambda$ is the mean
wavelength of X-ray photons in the dust frame, $\tau(\theta,\phi)$
is assumed to be small, and the light that multiply scatters into
the line of sight is ignored.

At first, for simplicity, we assume that $\tau(\theta,\phi)=\tau_0$
does not vary with $\theta$ and $\phi$ based on the fact that the
dust layer may have a small thickness (Vaughan et al. 2004, 2006).
For a bursting source, most of its energy is assumed to be emitted
in a narrow band on a time much shorter than $t_d$, and the flux is
then approximated as $F_u(t)=S_0\delta(t)$, where $\delta(t)$ is the
Dirac delta function and the source's trigger time is taken as the
time zero point. It follows from equation (9) that the total flux of
the echo is analytically expressed as
\begin{equation}
F_h(t)=\int I(\theta,\phi;t) {\rm cos}\,\theta{\rm d}\Omega={4\pi
S_0aD_s\tau_0j_1^2\{\hat{x}[\hat{\theta}(t)]\}\hat{\theta}(t)\over
\lambda D_{ds}\hat{x}[\hat{\theta}(t)]t}\,,
\end{equation}
where the function $\hat{\theta}(t)$ is defined as
$\hat{\theta}(t)\equiv[2ctD_{ds}/(1+z_d)D_dD_s]^{1/2}$, and
$\hat{x}(\theta)$ is defined as $\hat{x}(\theta)=2\pi
aD_s\theta/(\lambda D_{ds})$. Some formulae are given in the
Appendix for translating the Dirac delta function of $t$ into a
function of $\theta$.

This temporal behavior is illustrated in Figure 2 with the
dot-dashed line. The dust layer is assumed to be in the host galaxy,
and we have $D_d=D_s$, $z_d=1$, and $D_{ds}=100\, {\rm pc}$ in
Figure 2. Studies of interstellar extinction indicate that most of
the grains have a size near $a\sim0.1\,\mu {\rm m}$ and a
distribution in a wide range \citep{m77,d03}. Here we choose
$a=0.1\,\mu {\rm m}$ for a simple illustration. The X-rays in the
range $2--6\,{\rm keV}$ can be scattered most efficiently
\citep{m99}, and thus the average photon energy $\epsilon=6\, {\rm
keV}$ is used. The total X-ray fluence $S_0\sim10^{-6}\, {\rm
ergs\,cm^{-2}}$ in the energy band from about 2 to 10 keV is given.
The mean scattering optical depth $\tau_0=0.1$ is assumed in Figure
2, although $\tau_0$ is expected to be dependent on the photon
energy (e. g. M\'{e}sz\'{a}ros \& Gruzinov 2000; we account for this
below).

Note that in the X-ray scattering scenario, the size of the
interstellar dust grains is generally larger than the wavelength of
the X-rays. As very big particles block practically all light that
falls onto them, most of the scattered photons substantially
originate from diffraction at their edges. The diffraction of X-rays
through a dust layer causes an interferometric pattern in the
differential cross section as a function of the scattering angle, as
given in equation (8), only if equation (7) holds (e.g. Kr\"{u}gel
2003). This translates into the pattern of total flux versus time,
as shown in Figure 2. The significant semi-periodic interferometric
pattern indicated in these light curves reminds one of the X-ray
light curve observed in some long GRBs, e.g., GRB 050904
\citep{w06,c06}. Of course, this pattern may be smoothed out in some
other realistic situation, when there is a range of dust grain sizes
and the total flux is detected in a finite X-ray band (e.g. $0.3-10$
keV for XRT ). This is treated in some detail in the following
subsection.

The shallow decay phase shown in the light curve, which is common in
the small-angle scattering scenario predicted by equation (3), can
actually be attributed to the first maximum in the differential
cross section peaking at a very small scattering angle, with the
scaled notation $x=(2\pi a/\lambda)\alpha \simeq 2\pi a D_s \theta/
(\lambda D_{ds}) \simeq 1.5$ \citep{ah78}. The differential cross
section decreases dramatically at $x\gtrsim3.0$ \citep{ah78}, which
is translated into the decaying total flux at $t\gtrsim (1+z_d)D_d
D_{ds}\lambda^2/(8\pi c D_s a^2)\simeq 5\times 10^3s(\epsilon/6{\rm
keV})^{-2}(a/0.1 \mu{\rm m })^{-2}(D_{ds}/100 {\rm pc})$. After
that, a fast power-law decay of the maxima (roughly $\propto
t^{-2}$) emerges. Regardless, the small-angle approximation used
here (i.e. $\sin\alpha \sim \alpha$) is not violated throughout the
whole light curve, since the scattering angle $\alpha\simeq (c
t_d/D_{ds})^{1/2}$ is no more than several arcminutes at a delayed
time $t_d\simeq 10^5$ s, with $D_{ds}\simeq 100$ pc.

Furthermore, we also consider the scattering of a constant X-ray
beam emitted by the source within a time range $T$, i.e., the flux
is approximated as $F_u(t;t\leq T)=F_0$. The received flux can be
calculated by integrating equation (9) over $\theta$ and $\phi$. The
results with different $T$ are also shown in Figure 2. The on-going
flux $F_0=10^{-6}\,{\rm ergs\,cm^{-2}\,s^{-1}}$ is given, and the
other parameters are not changed. The light curves due to scattering
of a beam are similar to those of a pulse, except that an early rise
is expected in the former case.

\subsection{Numerical Treatment for a Practical Case}

Above we assume a  constant scattering optical depth $\tau_0$, a
grain size $a$, and a homochromous initial fluence $S_0$. In some
cases , $\tau$ may vary as the X-ray energy $\epsilon$ changes, and
the grain size may have a certain distribution. For the
Rayleigh-Gans approximation, the total scattering cross section
$\sigma_{sca}(\epsilon,a)\propto \epsilon^{-2} a^4$ is also inferred
(e.g. Mauche \& Gorenstein 1986), and it is suggested by
observations that $n(a)\propto a^{-q}$ \citep{m77}. Here, in our
treatment with only small scattering optical depth ($\tau \propto n
\sigma_{sca} $), we assume that
\begin{equation}
\tau(\epsilon,a)=A\left({\epsilon\over{1\,{\rm
keV}}}\right)^{-s}\left({a\over0.1\mu{\rm m}}\right)^{4-q}~~~~
{\rm for}~~~~\left\{\begin{array}{cc} a_-\lesssim a \lesssim a_+
\mu{\rm m} \\
 0.2\lesssim \epsilon \lesssim 10 {\rm keV}
\end{array}\right.,
\end{equation}
where $A=(5-q)\tau_{\rm keV}/[a(a/0.1\mu{\rm
m})^{4-q}]|_{a_-}^{a_+}$ is a constant in units of ${\rm cm}^{-1}$,
$\tau_{\rm keV}$ is the scattering optical depth at 1 keV, $q\simeq
3.5-4.5$, $a_-\simeq 0.005-0.025\,\mu$m, and $a_+\simeq
0.25-0.5\,\mu$m are inferred from observations \citep{m77,mg86,d03}.
Here $s\simeq 2$ is inferred from a single observation \citep{m90}
and then adopted by M\'{e}sz\'{a}ros \& Gruzinov (2000).

An initial source spectrum should also be taken into account. In the
soft X-ray band, this is given by the Band spectrum \citep{b93}
\begin{equation}
S(\epsilon)=B\left({\epsilon\over{100\,{\rm keV}}}\right)^\delta
{\rm exp}\left[-{(\delta+1)\epsilon\over E_p}\right],
\end{equation}
where $B$ is a parameter in units of ${\rm ergs}\,{\rm
cm}^{-2}\,{\rm keV}^{-1}$, and $\delta\simeq 0$ and $E_p\simeq 200$
keV are suggested by Preece et al. (2000). Thus, we can calculate
the measured flux, e.g., in the $0.3-10$ keV band by XRT.
Consequently,
\begin{eqnarray}
F_h(t)&=&\int\int {4\pi A B
aD_s\tau(\epsilon,a)j_1^2\{\hat{x}[\hat{\theta}(t)]\}\hat{\theta}(t)\over
\lambda
D_{ds}\hat{x}[\hat{\theta}(t)]t}\left({\epsilon\over{100\,{\rm
keV}}}\right)^\delta {\rm exp}\left[-{(\delta+1)\epsilon\over
E_p}\right]{\rm d}a{\rm d}\epsilon
\end{eqnarray}
Here the unabsorbed initial emission is still taken to be a pulse of
light, approximately as in equation (10). In the case of a beam of
light, the difference in the echo light curve is the short rising
time at the beginning time (e.g., see Fig. 2).

Equation (13) can be evaluated numerically with different parameters
considered. We find that the temporal behavior of $F_h$
significantly depends on three observable quantities of dust grains.
The first quantity is the position of dust $D_{ds}$. This is because
$t_d$ depends prominently on the position of the dust, which is
inferred from equation (1). The second quantity is the maximal size
of a dust grain, $a_+$, which is the large-size cutoff for the size
distribution. The third quantity is the index $s$ in the dependence
of dust-scattering optical depth on X-ray energy.

One prominent characteristic of this temporal behavior is a
distinguishable broken power law, with a break time dependent on the
parameters discussed above. As shown in Figure 3, a shallow decay
before the break and a slope $\simeq -2.0$ after the break are
clearly present in the light curves. The shallow decay before the
break is expected from our previous discussion and roughly
consistent with our estimates of small-angle scattering, shown by
equation (3), while the steep decay after the break can be
attributed to the quickly decreasing cross section of larger angle
scattering at later times. This feature of the flux ($\propto
t^{-2}$) is also implied in the previous discussion, e.g., as shown
in Figure 2, which is roughly consistent with the decreasing maxima.

Figure 4 plots the spectral evolution during dust scattering. In
general, the spectra soften as the flux decreases. The softening of
the spectra can be attributed physically to the diffraction effect
(treated as scattering here) which is described in the differential
cross section versus scattering angle (eq. [8]). Softer X-rays tend
to be scattered at a larger angle $\alpha$, with high-order maxima
in the differential cross section, and are thus received at a larger
angle $\theta$, which leads to a longer arrival time due to a longer
light distance (eq. [1]). Obviously, one can tell from Figure 4 that
the visible softening emerges at later time when the steep decay
($\propto t^{-2}$) in the light curves shows up. This corresponds to
a delayed time of $t_d\simeq (1+z_d)D_d D_{ds}\bar{\lambda}^2/(32
\pi c D_s \bar{a}^2)\simeq 3\times 10^4s(\bar{\epsilon}/1{\rm
keV})^{-2}(\bar{a}/0.1 \mu{\rm m })^{-2}(D_{ds}/100 {\rm pc})$,
where $h$ is the Planck constant, and $\bar{\lambda}$,
$\bar{\epsilon}$, and $\bar{a}$ are  the equivalent average photon
wavelength, energy, and dust radius, respectively.

\section{IMPLICATIONS FOR X-RAY AFTERGLOWS}

Above, we revisit the X-ray scattering off dust grains at GRB
stages. Now we apply these results to observational data and suggest
that some of the X-ray afterglows detected so far may be
alternatively explained as the emission from X-ray echoes, i.e.,
dust-scattering--driven afterglows.

1. A shallow decay followed by a ``normal" decay and a further
steepening is suggested by Zhang et al. (2006) to be characteristic
of almost all the $Swift$ GRBs. (1) In general, to account for
shallow decay, a continuous activity of the GRB progenitor is
expected \citep{dl98,zm01,d04} or a power law distribution of the
Lorentz factors in the ejecta is assumed \citep{rm98,sm00}.
Alternatively, we propose that this feature can be attributed to the
X-ray echo emission at early times. (2) To account for the steep
decay after the ``normal" decay, the relativistic jet effect has
been suggested \citep{r99,s99}. Alternatively, this feature can be
explained as the X-ray echo emission at late times. To summarize,
the features with a shallow decay followed by a ``normal" decay and
a further steepening are consistent with the X-ray echo emission
presented above, e. g., as shown in Figure 3.

Here, we apply equation (13) to two recently detected GRBs, 060813
and 060814, in Figure 5, where $\delta\simeq 0$ and $E_p\simeq 200$
keV are assumed. The consequent parameters are $a_+\simeq
0.5\,\mu{\rm m}$ and $D_{ds}\simeq 10$ pc for GRB 060813 and
$a_+\simeq 0.25\,\mu{\rm m}$ and $D_{ds}\simeq 30$ pc for GRB
060814. The other parameters, $a_-\simeq 0.025\,\mu{\rm m}$,
$q\simeq 4$, and $s\simeq 2$, are the same for the two GRBs.
Obviously, a shallow decay is common in early X-ray afterglows. This
favors the dust-scattering scenario, which predicts a shallow decay
at an early time, when the scattering angle is smaller (see also eq.
[3]). In any case, the detailed light curve depends on several
parameters that we mentioned in \S 3.2, and thus the early temporal
index varies in a wide range, e.g., $\sim [0,-1.0]$, and then
steepens into $\sim [-1.0, -1.5]$ at a moderate time (see also Fig.
3). Of course, as the scattering angle gets larger at later times, a
further steepening of the decay ($\propto t^{-2}$) is also predicted
in our calculations. Nevertheless, the steep decay ($\propto
t^{-2}$) is not observed in most X-ray afterglows \citep{s07,r06b}.
In the latter case, it may be due to a larger $D_{ds}$, which means
that an echo event with a small scattering angle will take place
with a longer duration, so that the shallow decay component will
last a longer time, thereby preventing the emergence of a steep
decay.

2. Most early X-ray light curves are found to decline rapidly in the
first few minutes, with a power-law index of $\sim 3$ or greater
\citep{t05,n06,z06}. In general, this feature can be taken as a GRB
tail emission arising from high angular latitudes \citep{kp00}. Then
a break to a shallower decay component, as mentioned above, commonly
shows up, which is defined as an X-ray hump by \cite{o06}. Early XRT
data reveal little evidence of spectral evolution across this
temporal break, but where evolution is seen, the spectrum tends to
get harder \citep{o06,n06}. Provided that the ``hump" component is
explained as the X-ray echo emission, this spectral feature is
basically consistent with our expectation. As shown in Figure 4, the
early-time spectrum of echo emission in soft X-rays would hardly
change from the initial prompt spectrum. Given an intrinsic
evolution of the prompt spectrum, the delayed echo spectrum can be
harder than the contemporaneous tail emission in the observer's
frame.

Our model predicts that a visible spectral evolution in XRT is
expected when the steep decay phase ($\propto t^{-2}$) emerges. It
is worth mentioning that the Rayleigh-Gans approximation is adopted
here in calculating the differential cross section. This
approximation works pretty well for normal interstellar dust and
energies at or above 2 keV. However, it overestimates the echo
emission at lower energies \citep{sd98}, where absorption of soft
X-rays is important. This absorption effect will weaken the
low-energy component in the spectra. Thus, the spectral shapes, as
shown in Figure 4, have been idealized in our treatment. Regardless,
a softening spectrum is expected when the steep decay phase
($\propto t^{-2}$) emerges. This feature needs to be verified by
time-resolved spectral analysis, in case the steep decay phase does
emerge.

3. Fluctuations or flares are observed in some GRBs (e.g. GRB
050904: Watson et al. 2006, Cusumano et al. 2006; XRF 050406: Romano
et al. 2006a; GRB 060713A: Guetta et al. 2006). Generally these
flares are thought to be caused by late internal shocks, similar to
those that produce the prompt emission (Zhang et al. 2006; Burrows
et al. 2005b; Fan \& Wei 2005; Wu et al. 2005; Perna et al. 2005;
Liang et al. 2006), or late external shocks \citep{p05b}. Both the
late internal shock model and late external shock model require
late-time activities of central engines \citep{w05}.

Alternatively, we have three reasons suggesting that these flares,
or some of them, might be due to dust-scattering:

1. Enormous flares are also observed in some GRBs (e.g. GRB 050502B:
Burrows et al. 2005a; Falcone et al. 2006), which requires a more
complicated theoretical explanation. However, these flares usually
begin with an excess absorbing column that softens as the flare
progresses. In addition, the afterglow intensity and slope are
similar before and after the flare \citep{b05a,f06}. These features
seem to favor the dust-scattering scenario, which does not disturb
anything when it turns off. The softening feature emerges when it
turns on.

2. Semi-periodic fluctuations are revealed in some GRBs (e.g. GRB
050904: Watson et al. 2006, Cusumano et al. 2006; GRB 050730:
Burrows et al. 2005b), which reminds one of the interferometric
pattern expected in the dust-scattering scenario under certain
assumptions. Although this feature may always be smoothed out in
practical cases (see $\S$ 3), one can see a hint of the physics this
may have unearthed.

3. Otherwise, if the dust layer is broken up or distorted by the GRB
progenitor or is fluffy, as in some cases with many holes and voids
\citep{w94,m95,pk96}, flares should also be expected in the echo
light curves. For example, a big jump in $\tau(\theta)$ around
$\theta_0$ causes a rapid rise and a rapid decay around $t_d\simeq
(1+z_d) D_d D_s\theta_0^2/(2c D_{ds})\simeq 10^3{\rm
s}((1+z_d)/2)(D_{ds}/100{\rm pc}) (\alpha_0/60'')^2$ due to the time
delay of the scattered flux introduced in equation (9), where
$\alpha_0$ is the angular scale from the GRB source of the
fluctuation of the dust intensity in the host galaxy. In any case,
we may be able to discover whether these flares are due to dust
scattering, after improving our model with specific dust details.

\section{CONCLUSIONS}

In this paper, we revisit the X-ray dust scattering in GRBs. First,
we give an estimate of the emerging flux during such an X-ray echo
event, which is expected to be dominant in X-ray afterglows. We find
that an initial pulse of X-rays from a normal GRB scattering off
dust grains in the host galaxy can be followed by a long-term
``afterglow" with almost the same amount of energy as expected in
the GRB X-ray afterglows. Second, we investigate the behaviors of
such an event, especially its light curves. We find that some of
these features are also consistent with observations of GRB X-ray
afterglows. We suggest that some of the X-ray afterglows from GRBs
(especially those without sharp rising and decaying flares) can be
understood in the dust-scattering--driven afterglow model. The
scattering of the prompt X-ray emission from GRBs off the host dusty
regions can be an alternative explanation for most of the features
observed recently in X-ray afterglows by {\em Swift} XRT.

Several properties of dust grains (e. g., $D_{ds},\, a_+$, and $s$)
are supposed to be relevant for temporal behaviors of X-ray echoes.
The most deterministic one is the position of dust (i.e. $D_{ds}$),
because it determines the time delay $t_d$ given in equation (1).
However, until now, we have not known this quantity very well.
Reichart (2001) assumed that there is a preburst, dense environment
due to the strong winds of GRB progenitors with an inner radius of
about several parsecs, while Madau et al. (2000) and Moran \&
Reichart (2005) suggested that this radius may be about 0.001--0.01
pc. M\'{e}sz\'{a}ros \& Gruzinov (2000) and Sazonov \& Sunyaev
(2003) assumed a GRB origin at the center of a uniform dusty region
with a radius of about 10--100 pc. In addition, the situation could
be more complicated, with the evolution of the dust grain population
considered. Waxman \& Draine (2000) suggested that dust grains will
be sublimated by the optical/UV flash of GRBs out to a distance of
about several parsecs (see also Perna \& Loeb 1998; Lazzati et al.
2001; Perna \& Lazzati 2002; Heng et al. 2007). Recently, Campana et
al. (2007) suggested a distance of several parsecs by analyzing the
evolution of the soft X-ray absorbing column around GRB 050904.
Here, we find that a $D_{ds}\sim$ of $\sim$ tens of parsecs is
consistent with {\em Swift} GRBs, based on our
dust-scattering--driven afterglow model.

It should be noted that X-ray echo emission is proposed here to be
due to dust scattering taking place at a distance of $\sim$tens of
parsecs from the GRB source. However, in the standard external-shock
model, relativistic shocks generally take place up to a distance of
$\sim 10^{17}$ cm. So technically this echo scenario does not rule
out the existence of emission from external shocks in both X-ray and
optical/NIR bands. In addition, there is indeed some evidence for
chromatic light-curve breaks, which may require that the X-ray and
optical emission have different origins \citep{fp06,pa06,r06b}.
Thus, there will be at least two types of X-ray afterglows if the
external-shock model and the dust-scattering model are both valid.

\acknowledgments  We are very grateful to two anonymous referees for
comprehensive comments, which have allowed us to improve the
manuscript. We would like to thank Binbin Zhang for providing the
X-ray afterglow data of GRB 060813 and GRB 060814. We also thank
Damin Wei, Yongfeng Huang, Xuefeng Wu, Yuanchuan Zou, Yunwei Yu, and
Xuewen Liu for helpful discussions, and Rosalba Perna and Bing Zhang
for valuable comments. This work was supported by the National
Natural Science Foundation of China (grants 10233010 and 10221001).
L.S. was also supported by the Scientific Research Foundation of
Graduate School of Nanjing University.

\appendix

\section{APPENDIX - FORMULAE FOR ANALYTICAL TREATMENT}

The Dirac delta function has the following properties:
\begin{equation}
\delta(-x)=\delta(x)
\end{equation}
\begin{equation}
\delta(ax)=|a|^{-1}\delta(x)
\end{equation}
\begin{equation}
\delta(x^2-a^2)=(2|x|)^{-1}[\delta(x+a)+\delta(x-a)]
\end{equation}

In $\S$ 3.1, $F_u(t-t_d)=S_0 \delta(t-t_d)$ is assumed for a GRB
pulse. Here, the delta function of $t$ can be translated into a
function of $\theta$:
\begin{eqnarray}
\delta[t-t_d(\theta)] &=&\delta[{(1+z_d)D_d D_s\over 2 c D_{ds}} \theta^2-t] \nonumber\\
&=&{2 c D_{ds}\over (1+z_d)D_d D_s}\delta[\theta^2-{2 c t D_{ds}\over (1+z_d) D_d D_s}] \nonumber\\
 &=&{c D_{ds}\over (1+z_d)D_d D_s \theta}
 \{\delta[\theta+\hat{\theta}(t)]+\delta[\theta-\hat{\theta}(t)]\}\nonumber\\
 &=&{c D_{ds}\over (1+z_d)D_d D_s \theta}\delta[\theta-\hat{\theta}(t)]
\end{eqnarray}
where the function $\hat{\theta}(t)$ is defined as
$\hat{\theta}(t)\equiv[2ctD_{ds}/(1+z_d)D_dD_s]^{1/2}$, and $\theta$
is always positive in our treatment.

\begin{figure}
\epsscale{.80} \plotone{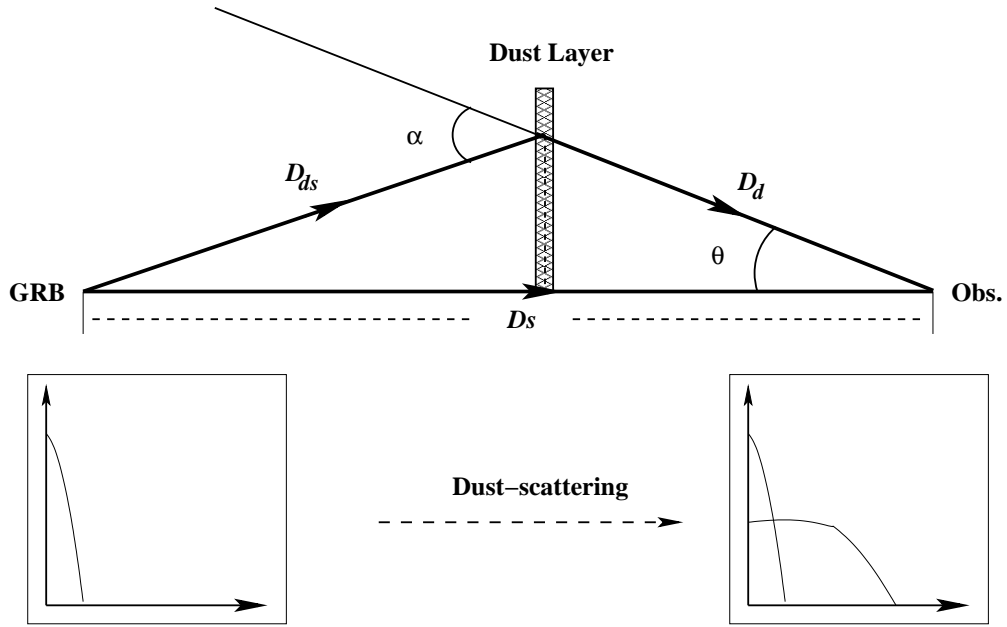} \caption{ Scheme of X-ray
small-angle scattering. The scattering angle is magnified to be
illustrative. In fact, the dust layer must be very close to the line
of sight. \label{fig1}}
\end{figure}

\begin{figure}
\epsscale{.80} \plotone{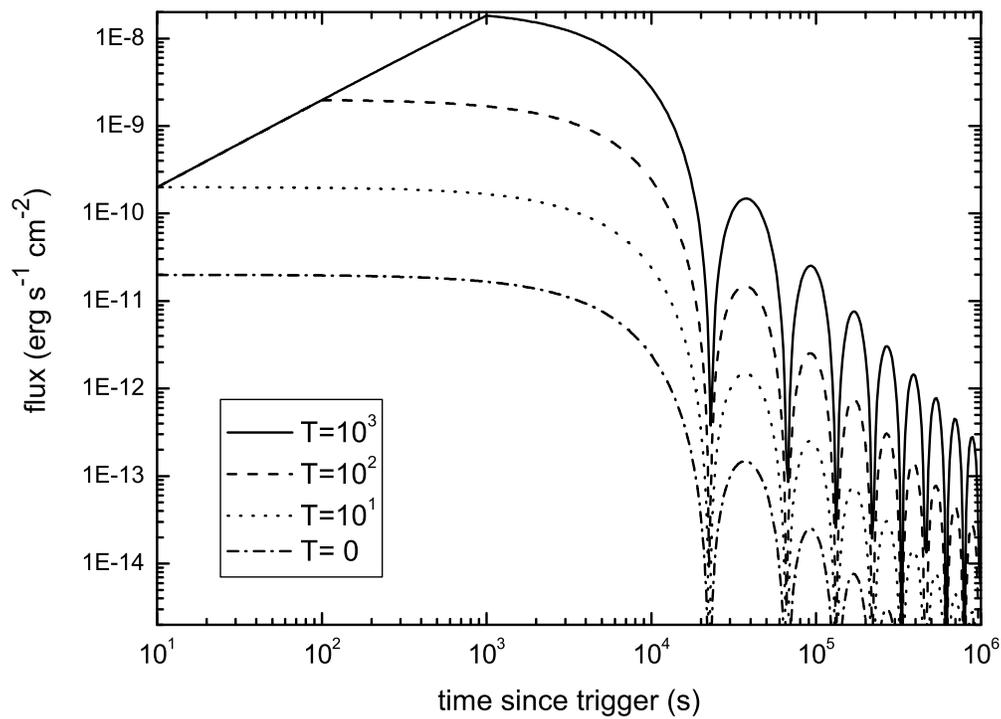} \caption{ Light curves of
dust-scattered X-ray echoes. $T$ is the timescale of the X-ray beam
defined in the text; the pure pulse is $T=0$. Parameters
$S_0=10^{-6} {\rm ergs\,cm^{-2}}$, $F_0=10^{-6}{\rm
ergs\,cm^{-2}\,s^{-1}}$, $D_s=D_d$, $z_d=1$, $D_{ds}=100{\rm pc}$,
$a=0.1\mu {\rm m}$, $\epsilon=6 {\rm keV}$, and $\tau_0=0.1$ are
assumed. \label{fig2}}
\end{figure}

\begin{figure}
\epsscale{.80} \plotone{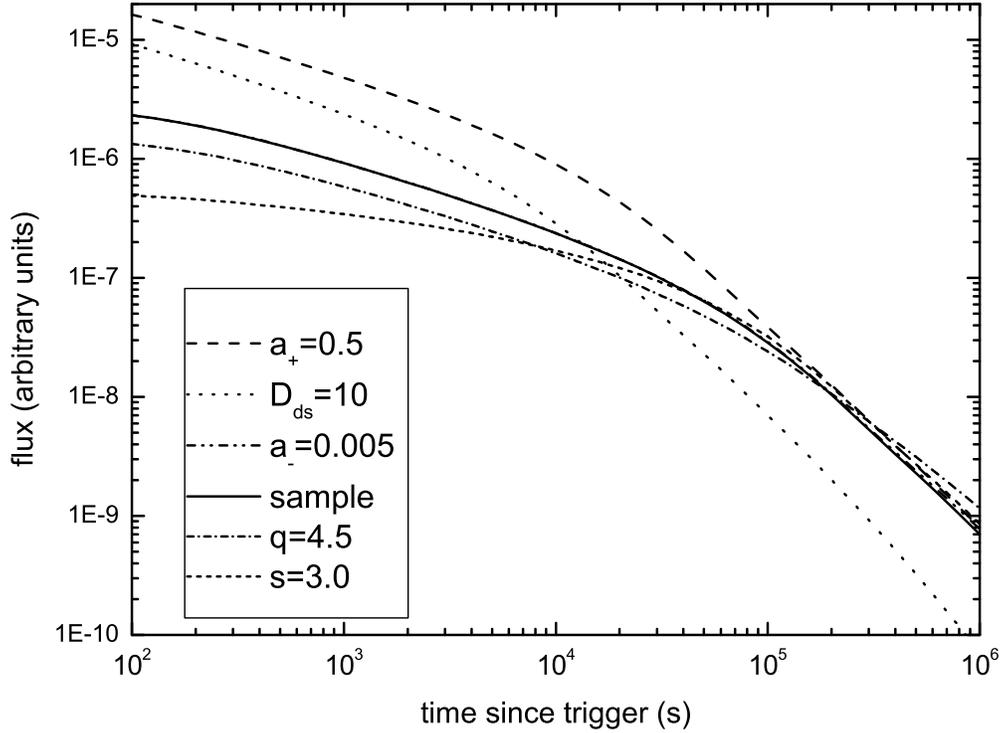} \caption{ Light curves of
dust-scattered X-ray echoes with detailed dust properties
considered. The solid line show the model with normal parameters
$a_-=0.025 \mu {\rm m}, a_+=0.25 \mu {\rm m}, q=3.5, D_{ds}=100 {\rm
pc}$ (we assume that the dust layer is in the host galaxy), and
$s=2.0$. Otherwise, we let only one parameter change in each line.
>From top to bottom, $a_+=0.5 \mu {\rm m}$, $D_{ds}=10 {\rm pc}$,
$a_-=0.005 \mu {\rm m}$, $q=4.5$, and $s=3.0$.\label{fig3}}
\end{figure}

\begin{figure}
\epsscale{.80} \plotone{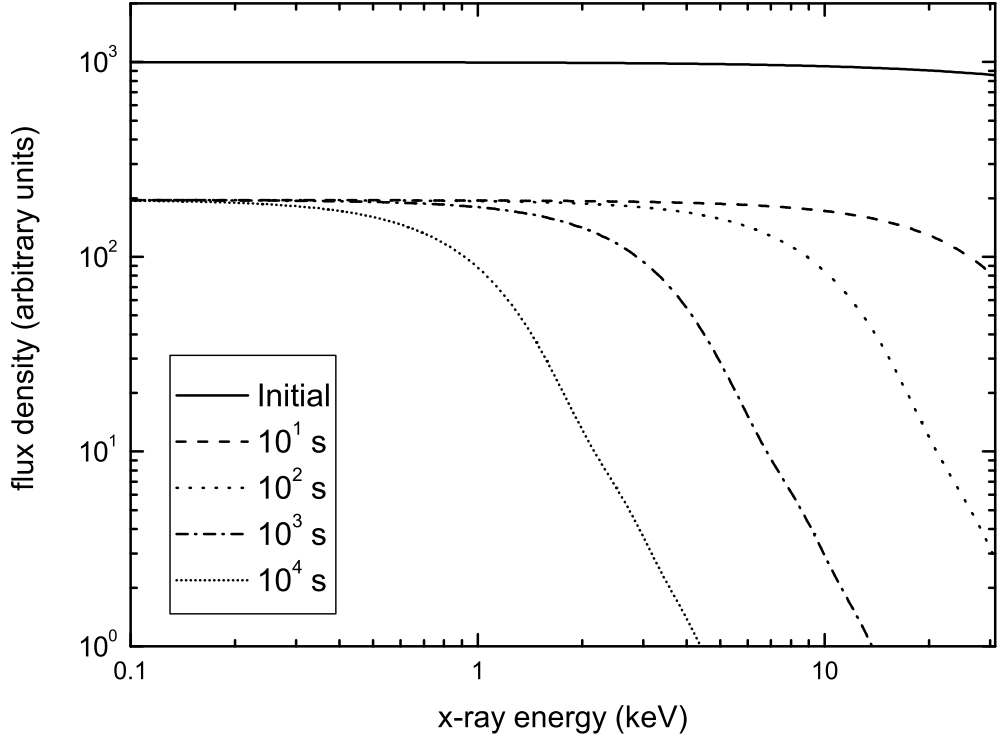} \caption{ Spectral evolution of
dust-scattered X-ray echoes. The solid line shows the initial Band
spectrum with $\delta=0$ and $E_p=200$ keV. The spectra of the
echoes at different times are shown from top to bottom. The dust
parameters are $a_-=0.025 \mu {\rm m}, a_+=0.25 \mu {\rm m}, q=3.5,
D_{ds}=100 {\rm pc}$, and $s=2.0$.\label{fig4}}
\end{figure}

\begin{figure}
   \plottwo{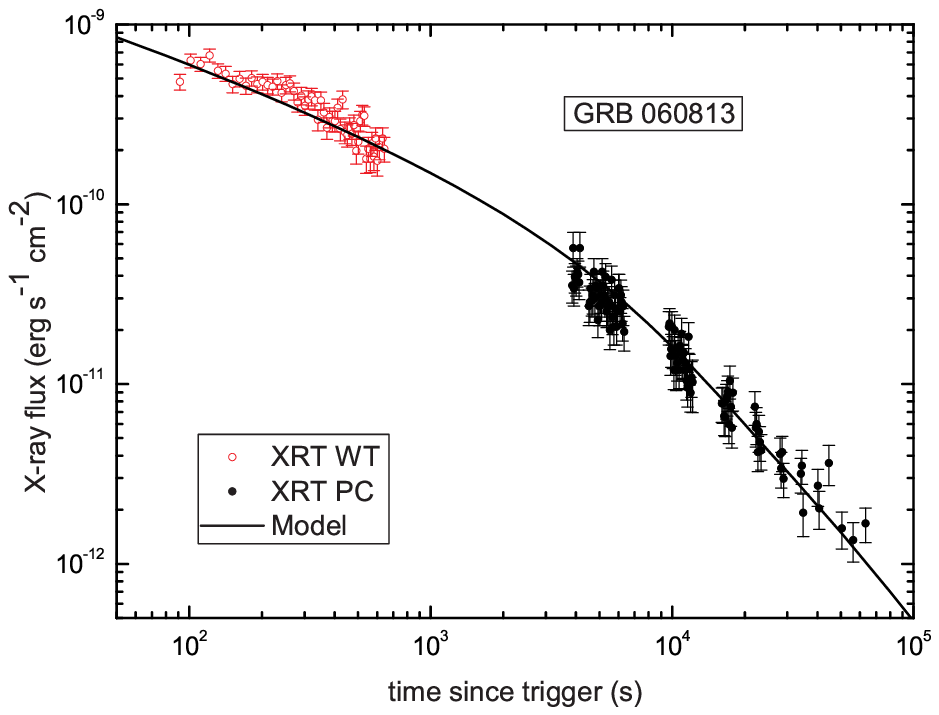} {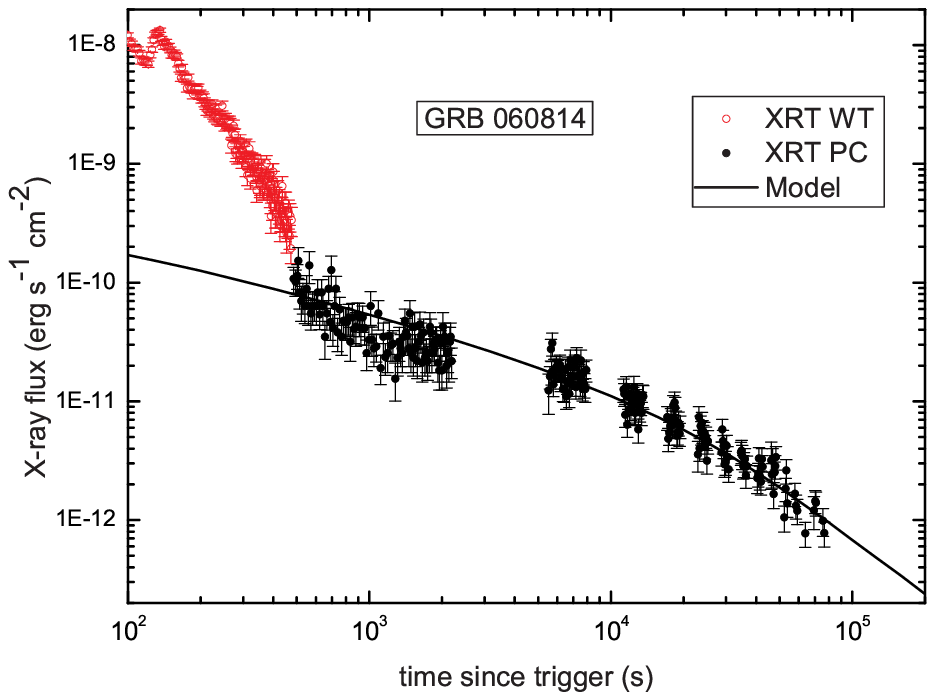}
   \caption{ X-ray afterglows of GRB 060813 and GRB 060814.
   $\delta\approx0$ and $E_p\approx200$ keV are
assumed. The consequent parameters are $a_+\approx0.5\,\mu{\rm m}$
and $D_{ds}\approx10$ pc for GRB 060813 and $a_+\approx0.25\,\mu{\rm
m}$ and $D_{ds}\approx30$ pc for GRB 060814. The other parameters,
$a_-\approx0.025\,\mu{\rm m}$, $q\approx4$, and $s\approx2$, are the
same for both of them. [{\em See the electronic edition of the
Journal for a color version of this figure.}] }
   \label{fig5}
\end{figure}

\end{document}